[1994/12/01]
\documentclass[draft]{article}

\usepackage{amsmath,amsthm}

\chardef\bslash=`\\ 

\theoremstyle{definition}

\theoremstyle{remark}

\newcommand{\eval}[2][\right]{\relax
  \ifx#1\right\relax \left.\fi#2#1\rvert}

\begin{document}

\title{ \Large \bf A Natural Concept of Image in the Physics of
 fr. Alonso de la Veracruz. \\}

\vskip.25in

\author{ Armando Barra\~n\'on.  \footnote{ Departamento de Ciencias B\'asicas, Universidad Aut\'onoma Metropolitana-A,
M\'exico, D.F. 02000}  }

\date{ December 10, 2000 }

\maketitle
\markboth{Natural Concept of Image in the Physics of fr. Alonso de la Veracruz.}
{ Natural Concept of Image in the Physics of fr. Alonso de la Veracruz.}
\renewcommand{\sectionmark}[1]{}

\begin{abstract}                

  Alonso de la Veracruz conducted a physical study of image, regarding the activities of Soul in which image is produced, as organic operations. This research was particularly important since, at that time, image was systematically used to propagate the European culture in New
 Spain. Besides, Alonso uses the visual radius to criticize the magical attitude popular in Renaissance, denying the far sight and fascinations abilities attributed to witches. Also, Alonso Guti\'errez applies the Aristotelian physics to deny the healing powers attributed to the kings of France and England, secularizing this way the gallic monarchies,   and several other superstitions of that time. This way, the physics of image developed by Alonso can be considered as close to the rationalism of 
Descartes inasmuch as Guti\'errez  criticizes the magical view of Renaissance and introduces geometrical elements to elucidate physical problems.

\end{abstract}

\section{Introduction}

   The $Studium$ $Generale$ at Tiripit\'{\i}o was founded in New Spain by the Augustinian order in 1540, as a school devoted to the Scholastic instruction of the monastic neophytes. In this academic institution fray Alonso, who received the sacred orders in Veracruz, served as Lecturer on Major Sciences and Theology, teaching the future Philosophy instructors of New Spain. At the
 convent of Tac\'ambaro, where Guti\'errez was named Provincial in 1545, Alonso taught Minor and Major Sciences to novices. His treatises on $Dialectic$ and $Physics$ were written in the period of time comprised between 1554 and 1557 \cite{1}. Alonso belongs to the group of rigid scholastics lead by his mentor Vitoria, influenced by the Renaissance and his works were written as draft notes for his lectures just as Aristotle did for the first time. Although Alonso shares with the rest of these Scholastic philosophers an abusive use
 of the $magister$ $dixit$ and a poor mastery of  argument, Guti\'errez introduces a Nature's conception related to his vibrant experience of the New World and important adaptations to comprise the Mesoamerican rationality.

  There were two general kinds of approach to knowledge for the Renaissance men. The first named pathetic, sought personal fusion with the Universe or God just like the mystic Cusano and the hermetic Ficino did. And the other was the distant movement which concentrated on the differences and effects between the subject and the object, having Leonardo and Valla as conspicuous exponents  \cite{2}. In this categorization, Alonso belongs to the distant attitude in 
as much as he finds a natural explanation for every event and only in  very extreme cases considers the influence of God.

 In this framework, Alonso wrote his $Physica$ $Speculatio$, dealing with the native system of beliefs and the magical conception peculiar to the Renaissance Age. As long as Guti\'errez introduces physical and geometrical elements to enlighten several superstitions, his contribution can be considered as a step forward into the Rationalism trend that lead to Modern Science. Moreover, this critical attitude to the royal magical power can be explained as a remnant of the Gregorian movement that secularized the European monarchies in the last stage of the Middle Ages.

\section{Physics and Soul in the Scholastic system.}

  In the Scholastic system the problems related to change, movement and \\
transformation of living things were considered as part of Physics \cite{vera1}  since it was devoted to obtain the forms or principles behind the  activities of the living things and corporeal substances. In this sense, the activities of Soul were regarded as a physical subject, as long as they are
 related to organic operations \cite{vera11}: `` tamen de intellectu, seu de anima intellectiva, in quantum forma est corporis, non creatur,
 nisi in corpore, et cum corpore, constituit rem naturalem: physici est consideratio". 

 Aristotle considers soul as the actuality of the body and an important element for living beings, rejecting the dualist approach of Plato and leaving the question of separability of soul and body as an open question ($De$ $Anima$ II 1, 412a17; II 2, 414a1-2). Aristotle regards perception as a definitional faculty for animal beings, fitted for sensible qualities instead of intelligible forms and therefore distinguished from mind ($De$ $Anima$ II 4, 415a20-21). 
This way Aristotle's hylomorphism, where perception is related to accidental forms and soul is an essential form, leads to the consideration of soul as a physical problem ($De$ $Sensu$ 1, 436b10-12).

  This is clearly stated by Averroes, who tried to purify Aristotle's physics of
any neoplatonic influence, in his $De$ $Anima$, 17th Commentary, where Soul
 is considered as a natural being: `` quia anima humana est de numero entium naturalium"  \cite{vera3} . Averroes regards the agent and material intellects as immortal and one for every men. Introduces a new corruptible `` passive intellect" \\(imagination) that acts on images ($phantasmata$) in order to obtain universal concepts and to impress them in the separated possible intellect.

 However, for the Aristotelian neoplatonic philosopher Temistius, named by his 
colleagues $eloquentiae$ $rex$, Aristotle's $De$ $Anima$ is a mathematical production lying between the natural Philosophy and Metaphysics: `` quod liber de anima est medius inter naturalem, \& metaphysicam scientiam: et sic mathematicus"  \cite{vera4}. Temistius considers that,  whereas separable from soul, agent and passive intellects belong to each person's soul, 
a perspective shared by Aquinas even when it contradicts the Aristotleian principle of passive mind corruptibility. The scholastic system considers active intellect ( $intellectus$ $agens$ or $nous$ {\it poi\^etikos}  ) as the source of the representations of all things since it imprints their intelligible impressions on the potential intelect ( {\it intellectus 
possibilis} or { \it nous dunatos} ). Temistius calls this action of the agent intellect on possible intellect as `` speculative intellect ", since at this moment a concept is formed in the human thought.

 And Alexander of Afrodisias, perhaps the most authoritative proselyte of Aristotle, considers the study of Soul as a part of  Natural Philosophy. In his view, Soul is corruptible since it is the form of the body. The material intellect, considered by him a passive intellectual faculty, resides  in the soul and dies with the body although the agent intellect lies outside of it. There is a material intellect for each person and only one divine agent intellect that is used by every one.

  Following Alonso de la Veracruz, Metaphysics is devoted to the study of isolated, immaterial, self sstaining  and incorporeal substances. Since the Soul does not fit into this class of substances, Alonso considers that it will not be studied by Metaphysics. For rational Soul attached to a body forms a natural body and is the first organic potential act received by life, namely of a man composed of natural elements, Alonso concludes that knowledge of Soul belongs 
to Natural Philosophy \cite{vera5}:  `` in quantum forma est corporis physici organici, ad naturalem philosophia pertinet directe."

\section{Image and Soul physics.}

   In the Sixteenth century, images ( $phantasmata$ ) are considered as a product of human Soul and, in regard to their relation with reality, as a physical problem. Image was also used by the Spaniards to solve the communicational problem involved in evangelizing. The Aristotelian 
conception on the rational control of the images produced by the brain was used by the Spaniards to explain the European cosmology. In New Spain this was 
made by the $tlacuilos$ of the Painting School at Tlatelolco, nearby the Colegio de Santiago, where Indians were instructed in reading, writing, music, latin, rethoric, logic, philosophy and native medicine   \cite{3}. 

   Malinalco convent frescos seem to be a product of this $tlacuilos$  \cite{4} since they trained the authors of these painting and supervised these productions \cite{5}  as written by chronicler Grijalva \cite{6}. They clearly show the fundamental symbols of solar Aztec religion and use them to translate the European paradise concept in terms of the shaman cultural complex  
\cite{7}.

   In the first stages mimic was used by friars until they attained a good knowledge of native languages. These theatre was substituted by the use of $lienzos$, namely pieces of fabric with drawings of the most important evangelic themes that were explained with the help of native translators  \cite{8} .

    This way the so called ``war of images", characteristic to the spiritual conquer of Mexico, was conducted on the basis of a cerebral function understood as organic and therefore subject to control of human action  \cite{9}. These ideological diffusion techniques were condensed in the important treatise $Rethorica$ $christiana$ written by fray Diego Valad\'es, son of a Spaniard 
conqueror possibly married with a tlaxcalteca woman \cite{10}. Even when Christendom always used images as an ideology diffusion instrument both in Byzantine and roman versions, friar Diego Valades  claimed Franciscan authorship in the use of these mnemonic $lienzos$ for indigenous instruction:

`` Though many others have made similar paintings (since it is not difficult to widen what once was invented) we are not looking for vulgar admiration as long as we never wrote such thing."

   European image assimilation by $tlacuilos$ is notorious in the graphical \\interpretations of the European engravings. Native and European symbols are mixed creating new imaginary entities, just as happened with the infernal $tameme$ found in Valad\'es $Rethorica$ $cristiana$. This image portrays an Indian lifter together with other demons  in the engraving $Tormentos$ $a$ $los$ $Pecadores$ (Sinners Torment) and was used to persuade neophytes of remaining in Christian purity.

\section{Physical causes and the wonderful things.}

   In his $Physicae$ $Speculatione$, fray Alonso considers natural events with a Renaissance approach, excelling the medieval conception where world was \\ separated in supralunar and sublunar, artificial and natural, natural and violent. Aristotelian natural conception considered instead the equation : $form=finality$, and gave an integral explanation of every phenomena as part of an integral world that Renaissance men considered as subject to modifications by human will  \cite{11}.
   Fray Alonso regards natural movement of heavens ignoring any intrinsic heaven Soul \cite{vera2} : 

`` In third place, considering whether the movement of heavens is natural or violent, and in Aristotle's words if heavens are not propelled therefore their movement is not natural. This follows in first place from the fact that every movement needs an intrinsic principle and in heaven the moving intelligence is not such a principle. In the second place, natural movement is neither regular nor uniform since it is faster in the final stage than in the beginning. 
And the contrary happens with violent movement. 
   Some others say that natural movement is still, such as every one sees that occur in the celestial movements. But the celestial movement is natural, therefore it's unnecessary to consider an intrinsic Soul in Heaven".

   Nevertheless, when the Aristotelian apparatus was incapable of explaining a wondrous event, Alonso admitted the miracle and criticized the Renaissance affection to magic and animism. Renaissance men considered themselves as magic  rulers of universe, following Pico de la Mirandola's human conception as the most wonderful part of Creation and `` concurrence of stable eternity and fluent time"  \cite{12}. This hermetic doctrine is clearly evoked in Hermes Trismegistus's $Asclepius$ \cite{13}:

``   Hence, Asclepio, men are a magnificent miracle and worth of reverence and honors. Since men reach the  Divine Nature as if they were Gods themselves, they are familiar with the demons since they share the same origin. Therefore, neglect in men the human side as long as they have recovered their hope in their divine nature. "

  Even Copernicus, well known for his mathematical approach and deductions, after showing his famous scheme that explains heliocentric theory, mentions the Hermetic conception of solar majesty \cite{14} : 

 `` Siquidem non inepte quidam lucernam  mundi, alii mentem, alii rectorem vocant.  Trimegistus visibilum deum.".

   Inquisitorial processes, commonly known as witch-hunt could never have happened without this kind of Rationalism that accepted the miracle or diabolic influence when the Aristotelian system failed. This attitude lead to a permanent hunt of native religions in the New Spain. Indian leaders, known as $caciques$ were considered diabolical and prosecuted by fray Juan de Zum\'arraga , who had already participated in Nueva Vizcaya, Spain, in a Witchcraft hunt and was invited to New Spain in recognition of these activities. \cite{15} 

   Fray Andr\'es de Olmos also considered native culture as diabolic and wrote a treatise on Demonology, explaining native rites, treating the encounters with the Indian priests as diabolic. As we know Indian priests used a disguise with the image of the very God they were revering  \cite{16}. And Mesoamerican codex represented these disguises, as Bodo Spranz has proved showing their fundamental characteristics in the Borgia Group Codex  \cite{17}. 
This costumes produced panic in Spaniards and Indians who faced the priests as dealing with supernatural entities.
\section{Veracrucian physics of image.}

    In the Aristotelian system, rational Soul understands the natural phenomena by means of images ($phantasmata$) which are a cerebral function and therefore material `` sed anima rationalis est huius modis; tum quia intelligere eius est per phanstamata" \cite{vera6}. And just
 because the vegetative and sensitive souls are corruptible forms, they must be of a material nature that must be studied by physics instead of mathematics or Metaphysics \cite{A}: `` cum tam vegetativa, quam sensitiva, sunt materiales formae, \& corruptibiles \& per se non possunt stare, \& educantur de potentia materiae" . Just because everything in the intellect comes from the senses, 
therefore intellect must use $phantasmata$, namely the organic-cerebral function in order to achieve thinking: `` quia nihil est in intellectu, quin prius fuerit in sensu: \& necesse est intelligentem phantasmata speculari: tamen intellecti vere perficitur, \& consumatur sine aliquo organo corporis" \cite{vera7}. 

   Since the sensible object is introduced in the senses as an image or \\representation, agent understanding uses $phantasmata$ to apprehend the being and everything circumventing him as an object:  `` quod cum accidens commune sit in phantasia, intellectus agens abstrahit a phantasmate speciem intelligibilem entis" \cite{vera8}. This way agent understanding knows the substance through proper accident since it manifests substance, as stated in Cayeto Teniense's $De$ $Anima$:   `` quibus intellectus possibilis praeparatur ad recipiendam speciem illius \\ substantiae, cuius sunt accidentia
 propria" \cite{vera10}. And the existence of the first engine is concluded from its regular and continuous movement for operation follows the being. Hence, separated substance is concluded from its accidental manifestation.
  In conclusion, accident is the principle of the knowledge of substance but not a principle of the substance being: `` Accidentia communia non sunt propia alicuius speciei, neque propria alicuius individui, ut constat: ergo non possunt nos ducere in cognitionem distinctam alicuius substantiae" \cite{vera9}.

   For Temistius sight depends on the eye organic conditions , the medium and the distance between the object and the eye namely the visual radius \cite{B}: `` quod radius visuales est species rei visibiles quae est corporalis pyramidis; cuius pyramidis basis, est in re visa: \& conus, in oculo vidente totum luminosum, pyramidem sui luiminis in quolibet puncto medij terminat". And the invisible is this way limited to the inability of the senses to see the object just like a very intense light, a diffuse image or a very weak light.  

   Guti\'errez recalls several opinions on the nature of hallucination. For instance Avicena considers hallucination as product of Soul imagination that can strongly influence in many ways the organs and even the external objects: `` Quo ad primum de fascinatione Avicenna 9. suae metaphysicae, \& 6. naturalium asserit esse, dicit supra corpus tantam habere virtutem, \& dominium : ut possit membra sola imaginatione diversi modi afficere" \cite{E}. And as told by the great physician Gentilis de Fulgino, witches can inflict damage on children because of their putrid humors `` quode vetuale ob humores putridos quos habent, \& corruptam complexionem, inficiunt teneros pueros" \cite{C}.  Also, Albertus Magnus attributes hallucination to the astrological configuration in which men are born: `` ut fascinationem ex constellatione coeleste evenire dicat: ut in nato in tali constellatione talis sunt fascinandi virtus" \cite{D}.
 Nevertheless, Alonso thinks that these ideas ignore God Will government on
 created and corruptible things (Avicenna, $De$ $las$ $Cosas$ $Naturales$, VI) and contradict the catholic religion. 

    Veracruz considers a mistake that Cornelius Agrippa, $haereticus$ \\ $pestilentissimus$, denies witchcraft in his $damnato$ book $De$ $Oculta$ $Philosophia$. In this book, Cornelius Agrippa mentions those magicians who, through \\mathematical learning, imitated natural productions such as `` the wooden dove of Architas, who did fly,and the statue of Mercury which did speak ". \cite{D1}

 For in erotic love, as Guti\'errez recollects, eyes can fascinate according to Plato's `` Phedro". And claims that a bishop Hyeronimus talks about some flowers that desiccate the eyes of those who see them \cite{G}.

  Veracruz concludes that bewitch is a product of Soul disposition and stars configurations, since many times people, trees or animals are bewitched and this is undeniable. Also choleric and melancholic individuals are easily bewitched . For some authors, wizards are fraudulent men with low ways and other consider them as capable of wondrous facts after a diabolic pact or because of a natural virtue, product of an astrological configuration. In the case of the medicine men it is necessary to be careful until being sure of what kind of individual they are : `` Quod si contingat ab eis aliquid mirandum fieri, vel virtute
 divinae tribuendum, vel (quod est magis verisimile) daemonis pacto sit" \cite{H}.

   Guti\'errez regards fabulous the bread baked to cure both rabies and the effects of the poison since this faculty can not be derived from the substance of the bread or the constellations:
`` Ex hoc sequitur fabulosum esse quod panis salutatus a salutatore sanet morsum rabidi canis, vel sit contra aliud venenum, quia haec virtus in pane non potest esse neque a complexione, neque a constellatione". Hence if some one is healthy because of this breads it will be either due to the diabolic participation or God's Will \cite{I}.

   Alonso considers the common belief on the healing powers of the kings of
 France and England as a mistake, since only God knows His Will. For the french kings this virtue was considered hereditary ``circa sensum tactus, an regibus Francorum  haec virtus iure haereditario sit" and the royal ring in  England was attributed with curing influences ``in regibus Angliae  benedicendi annulum, quo morbus contractionis  nervorum (qui calambre dicitur) sanatur" \cite{F}. 

   The Capeto lineage claimed a hereditary royalty sprung from Clodovean times just the same way as the English Normandy's dynasty alleged its origin in the Anglo-Saxon Ancestry.  This was part of a gallic tendency aiming to reach independence for the clergy and monarchy from Papal authority and leading to
 a Christian conception of the sacred nature of  royal power.

   The Royal Touch consisted in tracing the cross signal with the hands on the scrofula and afterwards washing the hands of the king. The water saved after this procedure was drunk by those sick for nine days and was believed to heal them. The fame of these kings attracted citizens from Spain, Italy and assured the inclusion of the Royal Touch in the medical treatises of the XIV century though until the XIV century this subject was fully considered in the medical literature. The famous philosopher William of Ockham considers that divine graces are transmitted in the royal unction which can be proved in the faculty of the kings to heal scrofula's patients.

    Since the XI century the Gregorian reform tried to secularize the royal power, exalting the spiritual faculties of the clergies to transubstantiate the wine and exorcise demons. This way, the Royal Touch was a theme proscribed from the ecclesiastical literature for about two centuries. The kings of Castilla 
were also considered as owners of the healing faculties, as stated in the $Speculum$ $Regum$ of \'Alvarez Pelayo, bishop of Portugal, where the King Sancho is told to have practiced an exorcism on a ``demoniaca" woman \cite{1J}. And Carlos of Viana, infant of Arag\'on and Navarra, was posthumously venerated in Poblet's Abby where his hand was used to heal the patients suffering of scrofula \cite {1aJ}. The emperor Carlos V used these miracle-working powers to legitimate the royal heredity in crisis after the Battle of Poitiers, invoking an apocryphal treatise attributed to Saint Thomas Aquinnas, where a fragment on the Royal Touch was included by his disciple fray Tolomeon after the death of this philosopher \cite{2J}. Therefore, Alonso rejects, on the basis of scientific operations, the sacred conception of the monarchy that arose as part of the gallic movement and adheres to the Gregorian attitude that secularized the monarchic power.

   Also, the oracular virtues attributed to some individual in certain days of the month, are clearly false since they would have this natural ability the rest of the month: `` Potest 
quidem esse, fateor, quod melius dispositus sit oculos uno die quam alio, \& noctu, quam diu tamen quod hoc sit in certis diebus solum, non apparet" \cite{J}.
  And the long distance sight is improbable only due to the demons influence, and all this because of the discussed physical constitution of sight. Though it's possible that some eyes could be able to see farther than the rest of the eyes: `` Quod possit esse talis dispositio in oculo alicuis hominis, qui videat ad magnam distantiam, ad quam alterius oculos no attingit, non est qui neget \cite{K}. But the ability to see the corpses buried in the churches, as some
respectable fellows ascertain, shold be result of diabolic influences or some other causes:
`` Et similiter cum intrant templum, ubi corpora mortuorum iacent, si submittant oculos, videant aperte viscera defunctorum [...] Possunt ista per daemonum illusiones fieri, tamen non damno" \cite{L}.  

\section{Conclusions.}

   In his $Physica$ $Speculatio$, dedicated to the Indians of the New Spain, Alonso de la Veracruz criticises the mystic approach of the hermetic natural philosophers and regards Soul as a physical subject, denying the existence of an intrinsic Soul responsible of the movement in Heavens. This way images and the invisible are considered as a physical problem that can be studied in terms of a geometric criterion provided by the visual radius. Since images are used by Soul to apprehend the being and everything circumventing him, they turn out to be an
 organic function subject to the control of human will.

   With this conceptual apparatus some superstitions, such as the healing powers of the kings related to the gallic movement that sought independence from the Pope, the magical powers attributed to the sorcerers in some days of the month, and the long distance sight, are easily discarded by Guti\'errez. Nevertheless,  any event unexplained by Aristotle's system, such as the continuous rains in Tenochtitl\'an (Mexico City), located in the $subtorrida$ zone, and the miraculous effects of some healing breads, should be therefore considered as a result of a miracle or a diabolic pact. This attitude supported  the inquisitorial actions taken against the Indian cultural system where hallucination \\ and fascination played important roles. 

Alonso's Physics 
of Image can be considered as a production that partially abandons the magic Renaissance approach and introduces mathematical and logical instruments to theorize  the real experiences provided by the discovery 
of the New World although accepting the possibility of miracle and diabolic influence. This way, Alonso's Physics of Image is closer to the epistemological rupture introduced by Descartes that lead to the modern science since it rejects the magical approach and applies a methodological and rational approach to study real events.



\end{document}